# A longitudinal analysis of students' motivational characteristics in introductory physics courses: Gender differences

Emily Marshman[1], Zeynep Y. Kalender[1], Christian Schunn[2], Timothy Nokes-Malach[2], and Chandralekha Singh[1]

*[1]Department of Physics and Astronomy, [2]Learning Research and Development Center*
*University of Pittsburgh, Pittsburgh PA 15260*

**Abstract:** The lack of diversity and the under-performance of underrepresented students in STEM courses have been the focus of researchers in the last decade. In particular, many hypotheses have been put forth for the reasons for the under-representation and under-performance of women in physics. Here, we present a framework for helping all students learn in science courses that takes into account four factors: 1) the characteristics of instruction and learning tools, 2) implementation of instruction and learning tools, 3) student characteristics, and 4) the students' environments. While there has been much research on factor 1 (characteristics of instruction and learning tools), there has been less focus on factor 2 (students' characteristics, and in particular, motivational factors). Here, we focus on the baseline characteristics of introductory physics students obtained from survey data to inform factor 2 of the framework. A longitudinal analysis of students' motivational characteristics in two-semester introductory physics courses was performed by administering pre- and post-surveys that evaluated students' self-efficacy, grit, fascination with physics, value associated with physics, intelligence mindset, and physics epistemology. We found that female students reported lower self-efficacy, fascination and value, and held a more "fixed" view of intelligence in the context of physics compared to male students. Female students' fascination and value decreased significantly more than males' after an introductory physics course sequence potentially signaling inequitable and non-inclusive learning environment. In addition, female students' view of physics intelligence became more "fixed" compared to male students' views by the end of an introductory physics course sequence. Grit was the only factor on which female students reported averages that were equal to or higher than male students throughout introductory physics courses. These gender differences can at least partly be attributed to the societal stereotypes and biases about who belongs in physics and can excel in it. These findings inform the framework and have implications for the development and implementation of effective pedagogies and learning tools to help all students learn.

## I. INTRODUCTION

Women are underrepresented in many STEM courses [1], and this is especially true in the field of physics [1]. Furthermore, prior research suggests that women generally under-perform in introductory physics courses [2-10]. Some efforts have been made to increase the diversity in physics courses [11-23], yet, the reasons for the low percentages and the under-performance of women in physics are not yet fully understood. Some researchers have hypothesized that the reasons may include, e.g., their prior preparation, career goals, self-efficacy, sense of belonging, mindset, and epistemology [24-63]. Increasing the diversity in the field of physics hinges, in part, on taking student characteristics into account in instructional design and implementation to improve teaching and learning for all students in physics courses.

In the past few decades, physics education researchers have investigated the challenges students face in learning physics and developed research-based instructional tools to improve student understanding of physics and their problem-solving and reasoning skills [64-76]. Indeed, effective pedagogies and learning tools build on students' prior knowledge and skills and provide appropriate coaching and support to help all students learn [77-81]. However, there has been less focus on how other factors such as aspects of students' motivation, students' environments, and the implementation of instructional strategies affect the efficacy of instructional interventions. Here, we first describe a holistic framework for engaged learning that not only takes into account students' cognitive skills but also incorporates their motivational beliefs (e.g., their self-efficacy, intelligence mindsets, and epistemologies), their environments, and the implementation of the pedagogies and learning tools to help all students learn physics. We then focus on the baseline measures of motivation of introductory physics students (in particular, women) obtained from survey data. This baseline data on students' motivation can aid in understanding the interaction between instructional interventions and aspects of motivation.

## II. BACKGROUND AND THEORETICAL FRAMEWORK INFORMING THE INVESTIGATION OF STUDENTS' MOTIVATIONAL CHARACTERISTICS

Below, we describe the theoretical framework, Strategies for Engaged Learning Frameworks (SELF) [82] and how prior research informs the framework. The Strategies for Engaged Learning Framework (SELF) (see Figure 1) is a holistic framework which suggests that instructional design and learning tools, their implementation, student characteristics, and social and environmental factors collectively determine how effectively students engage with and learn from instruction in a particular course. The framework consists of four quadrants as shown in Figure 1 and all of them must be considered holistically to help a diverse group of students learn effectively. The horizontal dimension involves the characteristics of learning tools (e.g., how the tool provides efficiency and innovation in learning and builds on students' prior knowledge) and students (e.g., the students' prior preparation and motivational characteristics), which both are taken into account when developing effective learning tools. The vertical dimension involves internal and external characteristics of the learning tools and the students. This dimension focuses on how the characteristics of the learning tools and students as well as the environments in which the tools are implemented are important to consider when helping students engage with and learn from instruction. The internal characteristics of the tool pertain to the tool itself (e.g., whether it includes formative assessment) and the external characteristics of the tool pertain to how the tool is implemented (e.g., whether the tool is framed appropriately to get student buy in). The internal characteristics of the students pertain to, e.g., their prior preparation, motivation, goals, and epistemological beliefs. The external characteristics of the students pertain to social and environmental factors such as support from mentors and balance of coursework. The SELF framework can be used as a guide in thinking holistically about students, learning tools, and the environment to inform instruction. Below we describe how the SELF framework is informed by cognitive theories and prior empirical research.

| | Learning Tool characteristics | Student Characteristics |
|---|---|---|
| **Internal Characteristics** | Factor 1. Learning tool characteristics (internal) – pertaining to features embedded in the learning tools that help students learn<br>• Based on "cognitive apprenticeship model" to promote mastery of material for a variety of students<br>• Include material providing scaffolding support<br>• Involve efficiency and innovation in learning<br>• Incorporate elements of productive engagement and productive struggle<br>• Involve formative assessment | Factor 2. Student characteristics (internal)<br>• Prior knowledge and skills<br>   o Prior preparation<br>   o Cognitive / metacognitive skills<br>• Motivational and affective factors<br>   o Goals<br>   o Interest and value<br>   o Self-efficacy, Identity<br>   o Epistemology beliefs<br>   o Intelligence mindset<br>   o Grit<br>• Self-regulation |
| **External Characteristics** | Factor 3. Learning tool characteristics (external) – pertaining to how the tool is implemented in a particular course<br>• Embed features to frame the importance of learning from tools and to get student buy in<br>• Embed motivational features within tools conducive to effective learning<br>• Reinforce learning by coupling learning of different students via creation of learning communities<br>• Make explicit connection between in-class lessons and out of class assignments and assessments<br>• Incentivize students to engage with tools via grades and other motivational factors<br>• Support to help students manage their time better<br>• Support to improve students' self-efficacy and epistemological beliefs | Factor 4. Student characteristics (external) - pertaining to the student-environment interaction<br>• Collaboration skills<br>• Balance of coursework and other work<br>• Family encouragement and support<br>• Support and mentoring from advisors and counselors<br>• Time-management |

**Figure 1.** Strategies for Engaged Learning Framework (SELF).

### A. Factors 1 and 2: Internal characteristics of learning tools and students

Factors 1 and 2 of the framework (the internal characteristics of the learning tool and students) are informed by several cognitive theories that point to the importance of knowing students' prior knowledge and difficulties in order to develop effective instructional tools. For example, Hammer proposed a "resource" model that suggests that students' prior knowledge, including their learning difficulties, should be used as a resource to help students learn better [78]. Similarly, the Piagetian model of learning emphasizes an "optimal mismatch" between what the student knows and is able to do and the instructional design [79]. In particular, this model focuses on the importance of knowing students' skill levels and reasoning difficulties and using this knowledge to design instruction to help them assimilate and accommodate new ideas and build a good knowledge structure. Vygotsky developed a theory which introduces the notion of the "zone of proximal development" (ZPD). The ZPD refers to the zone defined by the difference between what a student can do on his/her own and what a student can do with the help of an instructor who is familiar with his/her prior knowledge and skills [80]. Scaffolding is at the heart of this model and can be used to stretch students' learning beyond their current knowledge by carefully crafted instruction. These cognitive theories all point to the fact that one must determine the initial knowledge states of the students (i.e., students' prior knowledge and skills in Factor 2) in order to design effective instruction commensurate with students' current knowledge and skills (i.e., learning tool characteristics in Factor 1).

For example, instructional design that conforms to the field tested cognitive apprenticeship model [81] can help students learn effectively (see Factor 1). The cognitive apprenticeship model involves three major components: modeling, coaching and scaffolding, and weaning. This approach has also been found effective in helping students learn effective problem-solving heuristics and develop their cognitive skills. In this approach, "modeling" means that the instructor demonstrates and exemplifies the skills that students should learn. "Coaching and scaffolding" refer to providing students suitable practice, guidance, and feedback so that they learn the skills necessary for good performance. "Weaning" means gradually fading the support and feedback with a focus on helping students develop self-reliance. Much research in physics education has focused on these cognitive factors in developing effective instruction and assessment. For example, in physics, tutorials, peer instruction (clicker questions with peer discussion), collaborative group problem solving with context-rich physics problems, POGIL (process-oriented guided-inquiry learning) activities, etc. have been found effective in helping students learn [64-76]. Furthermore, several validated conceptual standardized surveys have been developed [83] to assess students' conceptual understanding of various physics topics.

However, Factors 1 and 2 are also informed by several "non-cognitive" aspects such as students' motivational characteristics, affective factors, and self-regulation. Several studies have focused on students' motivation (e.g., factors such as fascination and value associated with physics, self-efficacy, grit, intelligence mindset, and physics epistemology) [25-63]. For example, researchers have focused on students' interest in and value of science and their relationship to performance and STEM degree achievement [25-35]. Research suggests that interest in math and science is associated with the number of math and science courses taken in high school and career aspirations [28]. In addition, self-efficacy is another factor that can impact students' motivation and learning [36-46]. Self-efficacy is the belief in one's capability to be successful in a particular task, subject area, or course [36]. Students with high self-efficacy often exhibited effective learning behaviors during conceptual learning, such as self-monitoring and persistence, were less likely to reject correct hypotheses prematurely, and were better at solving conceptual problems than students with low self-efficacy but of equal ability [41]. Self-efficacy has been shown to positively predict performance in science courses and scores on the Force Concept Inventory [16]. Moreover, in Peer Instruction [69] environments in a physics course, students with low self-efficacy are more likely to switch to a wrong response after discussions with a peer than those with high self-efficacy [44]. Self-efficacy is also positively correlated with expected grades in physics [40], enrollment in STEM courses, and career choices in STEM [see ref. 27 and references therein].

Another factor that is associated with motivation is the construct of grit, introduced by Duckworth et al [47-52]. They defined grit as a trait-level perseverance and passion for long-term goals—a capacity to sustain both effort and interest in projects that take months or even longer to complete and adherence to goals even in the absence of positive feedback. Grit has been shown to predict achievement in challenging domains over and beyond measures of talent [47]. However, to our knowledge, there are no studies of students' grit within an introductory physics course sequence.

Dweck and colleagues defined another motivational construct that can affect motivation and learning, i.e, theories of intelligence (or intelligence mindset) [53-57]. Theories of intelligence involve views about the nature of intelligence—an "entity theory" in which intelligence is viewed as a fixed trait that one is born with or an "incremental theory" in which intelligence is viewed as malleable and can be shaped by the environment. Research has shown that children's implicit theories about the nature of intelligence impacts the goals they pursue, their response to difficulty, and how well they do in school [53]. Students who view intelligence as a fixed trait are concerned with demonstrating intelligence and prefer tasks in which they feel capable and smart. When faced with difficult tasks, students who view intelligence as a fixed trait tend to become debilitated and disengage. On the other hand, students who view intelligence as malleable are concerned with learning new concepts and improving competence. When faced with difficult tasks, these students appear to experience less anxiety, put forth more effort, and increase their engagement [53].

Students' epistemologies, or their beliefs about knowledge and learning, can also affect their motivation and learning of science [59-63]. Students' epistemologies in physics includes beliefs about the nature of physics (the belief that that physics knowledge is a set of disconnected facts and formulas or an organized system of interconnected concepts and principles), agency in learning physics (the belief that physics knowledge comes from authority as opposed to constructing their own knowledge), and study strategies (the belief that one should memorize formulas as opposed to understand them). Hammer's study on college students showed that students' ideas about knowledge and learning in physics affected how they solved homework problems during think-aloud interviews [58]. Physics education researchers have developed surveys that assess students' epistemologies in physics [59,84]. Research has found that students' responses to epistemological surveys stay approximately the same or become less "expert like" after taking traditionally taught introductory physics courses, e.g., students view physics knowledge as a collection of formulas as opposed to a set of interconnected concepts and view knowledge as coming from authority as opposed to developing their own understanding [59].

### B. Factors 3 and 4: External characteristics of learning tools and students

We note that instructional tools and student characteristics (Factors 1 and 2) do not exist in a "vacuum." One must also take into account the interplay between instruction, students, and the environment. Factors 3 and 4 focus on the environment, i.e., how instructional design is implemented in a particular course and students' social environments, respectively. One example of Factor 3 involves "framing" of instruction to achieve student "buy-in," which is helpful in motivating students to engage with learning. Several studies have shown that providing rationales that, e.g., identify why it is worth the effort to engage and communicates why it is useful to students can help them engage constructively with a lesson [85,86]. Motivational researchers also posit that providing stimulating and interesting tasks that are personally meaningful, interesting, relevant, and/or useful to students can increase their interest and value in a subject, increasing their motivation and engagement in learning [87]. In physics, researchers have developed "context-rich" physics problems, i.e., problems that involve "real-world" applications of physics principles and are complex and ill-defined [88]. These types of problems can often increase students' interest and value associated with physics. Furthermore, instruction that fosters a community of learners can also encourage productive engagement in learning. Within this community of learners, students are encouraged to construct their own knowledge and are held accountable to others, which can, in part, encourage them to engage deeply with the content [89,90]. Factor 4 focuses on the student-environment interaction. Indeed, having supportive parents, teachers, and mentors can be beneficial in fostering students' motivation and engagement in the classroom [91]. Students' time management skills have also been shown to correlate with performance in college [92].

In sum, the four factors of the SELF Framework can be thought of holistically when designing instruction. We note that each of the factors can interact with other factors. For example, the way in which instructional tools are implemented (Factor 3) is informed by the characteristics of the learning tool (Factor 1), the characteristics of the students (Factor 2), and the way the students interact with their social environments (Factor 4). Furthermore, the student characteristics in Factor 2 can inform the learning tool characteristics (Factor 2), the implementation of learning tools (Factor 3), as well as how the student interacts with the environment (Factor 4). Here, we investigate baseline measures of, e.g., student characteristics in Factor 2 such that instructors, instructional designers, and researchers can develop effective learning tools and implement them appropriately to help all students learn better.

### C. Gender differences in Factor 2 of the SELF framework

In this study, we focus on baseline measures of student characteristics in Factor 2, and in particular, we focus on gender differences in motivational characteristics throughout an introductory physics course sequence. Prior research has already revealed several gender differences in regards to Factor 2 of the SELF framework. For example, on many physics conceptual tests, female students underperform male students [2-8,83]. Doctor and Heller found a consistent gap in the scores of men and women on the Force Concept Inventory in introductory undergraduate physics courses, with men scoring, on average, approximately 13% higher than women on the posttest [7]. A prior study has also showed that women had lower gains on the Force Concept Inventory than men, despite the fact that there were no differences in course grades [6]. Similarly, women score lower than men on other conceptual surveys, such as the Test of understanding Graphics in Kinematics and Determining and Interpreting Resistive Electric Circuit Concepts Tests [83].

Researchers have also discovered gender differences with regard to motivational characteristics (see Factor 2 of the SELF framework). For example, one of the reasons women choose not to pursue STEM degrees is because their career interests focus on altruistic efforts and helping others, and they did not view STEM fields as those in which they could help others [30]. However, one study has shown that when a STEM career was presented to females as beneficial to society, their interest in the field increased [30]. Furthermore, females demonstrate lower self-efficacy relative to males beginning in middle school and throughout high school and college [see ref. 27 and references therein]. Females' intelligence mindset has also been investigated. It was found that girls who attributed intelligence to effort and learning had math grades comparable to those of their male classmates and superior to girls who viewed intelligence as a fixed trait [54]. Research has also shown that individuals targeted by ability stereotypes (e.g., underrepresented students in STEM courses) tend to show similar characteristics of individuals who believe that intelligence is fixed—they tend to choose easier, success-assuring tasks when their abilities are subject to scrutiny or if their ethnicity or gender is made more salient [55], experience anxiety when the tasks are evaluative and challenging [56], and devalue ability domains in which they have performed poorly [57]. Additional research has also shown that girls who viewed math ability as a trait (had a fixed view of intelligence) and also experienced stereotype threat (that girls are not good at math) had decreased motivation and interest in pursuing math careers [53]. However, to our knowledge, there have been relatively few studies of students' theories of intelligence in physics courses [76]. On the Colorado Learning Attitudes about Science Survey (CLASS) [84], which is a survey of students' epistemological views about physics, women are generally less "expert-like" on statements involving real-world connections, personal interest, problem solving confidence, and problem solving sophistication than males.

### D. Longitudinal studies of factors related to students' motivation

Here we examine male and female students' motivational characteristics longitudinally, i.e., over the course an introductory physics sequence. We note that there have been relatively few longitudinal studies of students' motivation in physics [16,93-98]. In the study by Bates et al. [93], researchers used the Colorado Learning Attitudes about Science Survey (CLASS) to investigate how undergraduate physics students' attitudes changed over a period of three years. They found that "expert-like thinking" was unchanged over the duration of the undergraduate physics program. Gire et al. [96] also performed a similar study in which the CLASS was administered to students in undergraduate and graduate physics courses over a period of two years. They found that physics majors have relatively "expert-like" views when they begin their undergraduate study of physics and maintain these views throughout the undergraduate physics program. Cavallo et al. [16] investigated gender differences in learning constructs and shifts in a yearlong, inquiry-based physics course for life science majors. They found that males had significantly higher self-efficacy, performance goals, and physics understanding compared to females, and these differences persisted throughout the course. In addition, they found that self-efficacy significantly predicted physics understanding and course achievement for both men and women. Lindstrom and Sharma [97] investigated students' self-efficacy in a first-year university physics course. They found that women consistently reported lower self-efficacy than males. Both gender and prior formal physics instruction impacted students' self-efficacy.

However, none of these studies focused on students' fascination with physics, value associated with physics, grit, or intelligence mindset—all of which may play a role in students' motivation and learning and need to be taken into account holistically as posited by the SELF framework. Furthermore, to our knowledge, there have been no longitudinal studies in physics that focus on factors related to females' motivation. Thus, we investigated student characteristics (factor 2), i.e., fascination with physics, value associated with physics, self-efficacy, grit, intelligence mindset, and physics epistemology. We report baseline data on these measures for both men and women in physics and how their motivational characteristics change over the course of an introductory physics sequence. The baseline data can be a stepping stone for developing better learning tools (factor 1) and determining more effective ways to implement learning tools (factor 3) in order to increase student engagement and success in physics courses. Providing a baseline can also help determine for which motivational factors there are gender differences and when those differences are present (e.g., at the beginning of a physics course sequence or later). This information can inform future instruction and interventions. For example, if there are differences between male and female students' interest at the beginning of a physics course sequence, one intervention may involve boosting female students' interest in physics early on in the course.

## III.   METHODOLOGY

To investigate students' fascination with physics, perceived value of physics, physics self-efficacy, grit, intelligence mindset, and physics epistemology, we created a 29-item questionnaire in which we adapted items from previously published surveys for each construct [48,53,59,84,99-101]. We administered the survey in introductory physics courses three times over a period of one academic year. Below, we describe the validation and administration of the survey as well as how the student demographics (such as gender and race) were linked to students' responses on the surveys.

## A. Development of the survey

The survey on motivational factors was developed by four of the researchers iteratively. The researchers first decided which constructs to assess based upon a review of the literature on student motivation (see the background section). Then, initial versions of the survey were iterated which included a set of questions from other well-validated surveys in the literature on motivation [48,53,59,84,99-101]. Surveys that served as sources were selected based on prior empirical validation and appropriateness for university undergraduate students. Individual items were selected to cover the conceptual space of each construct without redundancy; surveys sometimes improve reliability by including nearly identical questions. Minimalistic scales for each construct were necessary because many constructs were included in the overall survey, and the researchers decided that the students must be able to complete the survey in fifteen minutes or less to encourage high completion rates and longitudinal use. The final version of the survey includes 29 Likert-scale items. All items were on a Likert scale of 1-4 except for grit, which was on a scale of 1-5. Table I shows a representative list of items in survey, the corresponding motivational or epistemological factor, and the original survey the item comes from. The final version of the survey is included in the appendix.

**Table I.** For each factor, number of items, Cronbach alpha from each of the first two administrations ($\alpha_1$ and $\alpha_2$), example survey items, and source of items.

| Factor | Example Survey Item | Scale | Original survey |
|---|---|---|---|
| Fascination with Physics<br>3 items<br>$\alpha_1$=0.63, $\alpha_2$=0.67 | • I wonder about how nature works | Never<br>Once a month<br>Once a week<br>Every day | see ref. [99] |
| | • In general, I find physics | Very boring<br>boring<br>Interesting<br>Very interesting | |
| Valuing Physics<br>5 items<br>$\alpha_1$=0.69, $\alpha_2$=0.70 | • Knowing physics is important for being a good citizen …<br>• Physics makes the world a better place to live … | No!<br>no<br>yes<br>Yes! | see ref. [99] |
| Physics Self-Efficacy<br>6 items<br>$\alpha_1$=0.75, $\alpha_2$=0.74 | • I am often able to help my classmates with physics in the laboratory or in recitation …<br>• If I study, I will do well on a physics test … | No!<br>no<br>yes<br>Yes! | see refs. [84,99-101] |
| Intelligence Mindset<br>4 items<br>$\alpha_1$=0.65, $\alpha_2$=0.64 | • You have a certain amount of intelligence, and you can't really do much to change it ….<br>• Anyone can become good at solving physics problems through hard work … | Strongly disagree<br>Disagree<br>Agree<br>strongly agree | see refs. [53,59] |
| Grit<br>4/3 items<br>$\alpha_1$=0.58, $\alpha_2$=0.68 | • I often set a goal but later choose to pursue a different one ….<br>• I finish whatever I begin ….<br>• I have difficulty maintaining my focus on projects that take more than a few months to complete. | Not like me at all<br>Not much like me<br>Somewhat like me<br>Mostly like me<br>Very much like me | see ref. [48] |
| Physics Epistemology<br>8 items<br>$\alpha_1$=0.64, $\alpha_2$=0.70 | • I do not expect to understand physics equations in an intuitive sense; they must just be taken as givens.<br>• When doing an experiment, I try to understand how the experimental setup works …. | Strongly disagree<br>Disagree<br>Agree<br>Strongly agree | see refs. [59,84] |

For each survey item, students were given a score of 1-4 (or 5, for the questions related to grit). For the items related to fascination, value, grit, and self-efficacy, a high score means that a student is highly fascinated by physics, values physics highly, and has a high level of grit and self-efficacy. Furthermore, for the factor of intelligence mindset, a high score means that a student has a malleable view of intelligence, whereas a low score means that a student views intelligence as a fixed ability. For the factor of physics epistemology, a high score means that a student has a more "expert-like" view of knowledge and learning in physics, whereas a low score means that a student has a "novice-like" view. We note that some of the items on the survey were reverse coded since the statement was posed in a negative way (e.g., a student who strongly agrees with the statement "I do not expect to understand physics equations in an intuitive sense; they must just be taken as givens" would be given a score of 1). For each factor (e.g., fascination, value, grit, self-efficacy, intelligence mindset, and physics epistemology), each student was given an average score. For example, a student who answered "Yes!" to two of the fascination questions and "no" to the other fascination question would have an average fascination score of (4+4+2)/(3 total questions) = 3.33.

### B.  Administration of the survey

The survey was administered to students in algebra-based and calculus-based physics courses at the beginning of the fall semester of 2015, the beginning of the spring semester 2016, and the end of the spring semester 2016. The physics courses were varied in the type of pedagogy used, but the majority of the courses were lecture-based physics courses. Physics 1 courses included topics such as kinematics, forces, energy and work, rotational motion, gravitation, and oscillations and waves. Physics 2 courses included topics such as electricity and magnetism, electromagnetic waves, images, interference, and diffraction. The algebra-based physics courses are typically taken by students majoring in many different areas but planning on pursuing medicine or other health professions, for which the two-semester physics sequence is required. The calculus-based physics courses are typically taken by engineering and physical science majors as a required course for those majors. The physics courses included three lecture hours and one recitation hour per week. The recitations were mandatory and included a weekly, low stakes quiz. Most of the physics courses were in a traditional, lecture-based format, but there was one calculus-based and one algebra-based course that was in the "flipped course" format. In the "flipped" courses, students watched lecture videos before attending the lectures. In the lectures, they worked on collaborative group problem solving and clicker questions. Thus, some of the courses involved in this study involved characteristics from Factors 1, 2, and 3 in the SELF framework, i.e., providing scaffolding support, involving formative assessment, taking into account students' prior knowledge and skills, and use of collaborative learning and grade incentives to encourage students to engage with learning tools. However, we note that most of the courses did not explicitly take into account students' motivational characteristics in factor 2 and did not necessarily include features from factor 3 such as framing the importance of learning from tools and to get student buy in and embedding motivational features within tools that are conducive to effective learning.

The survey was typically administered in the first and last recitations of the course, although some instructors chose to give the survey in the lecture portion of the course. Furthermore, in the first round of administration (Fall semester 2015), some instructors chose to give the survey in a written format, whereas others chose to give the survey in an online format outside of class. In the first round of administration (Fall semester 2015), we found that the participation rates were significantly lower for students who were given the online format of the survey. Thus, in subsequent administrations of the survey (Spring semester 2016), we asked instructors to administer the survey in a written format. The survey was completed by most students in about 10-15 minutes. In general, there were no grade incentives given to students who took the survey. The instructor or teaching assistant responsible for giving the survey was given the following script to announce before administering the survey to the students: "We are surveying you on your beliefs about physics in order to improve the class. Your responses will not be evaluated for grades except to make sure the responses were done seriously, rather than randomly." This script encouraged students to take the survey seriously.

### C.  Validation of the survey

At the beginning of fall 2015 and the beginning of spring 2016, we analyzed the internal consistency of the subscales, i.e., fascination, value, self-efficacy, intelligence mindset, grit, and physics epistemology. After the initial reliability of the subscales in the fall 2015, we removed one of the statements in the grit subscale ("Setbacks don't discourage me") since the wording may have been confusing for students and it did not correlate well with other statements related to grit. All alphas are above 0.60 across both time points (see Table I) which is considered fairly good, especially since some of the scales have only three items. The scales with five or more items all had a Cronbach's alpha of 0.70 or higher. No substantial increases in alpha for any of the scales could have been achieved by eliminating items.

To establish the separability of six different subscales along with validity of items as clear indicators of the scale to which they were assigned, we performed an exploratory factor analysis on the items in the survey based upon the data at the beginning of the spring 2016. A principal components analysis with Varimax rotation method was used, and the initial eigenvalues indicated that the first six components (all with eigenvalues greater than 1) explained a total of 49% of the variance (the 7th component explained only an additional 3.6% of the variance). Item loadings on each component are given in the appendix. Component 1 is related to self-efficacy, component 2 is related to value associated with physics, component 3 is related to physics epistemology, component 4 is related to fascination with physics, component 5 is related to intelligence mindset, and component 6 is related to grit. Thus, the data supported the existence of six separable scales, and items loaded on the scales as intended.

We hypothesized that males and females may respond differently to the intelligence mindset items on the survey. For example, two of the items in the survey related to "general" views about intelligence mindset: "You have a certain amount of intelligence, and you can't really do much to change it" and "No matter how much intelligence you have, you can always change it quite a bit." However, the other two items related to intelligence mindset in the survey were embedded in a physics context: "Anyone can become good at solving physics problems through hard work" and "Only very few specially qualified people are capable of really understanding physics." We postulated that males and females may answer questions related to "physics" intelligence mindset differently due to the stereotype that men perform better in logical, math-intensive fields and women perform better in communication and writing-intensive fields [53]. Thus, in the results section, we report students' average "general" intelligence mindset and "physics" intelligence mindset separately.

### D.  Connecting student demographics to the survey

To connect student demographics such as gender and race to students' responses on the survey, we collected data on students who were enrolled in physics courses from the University's data warehouse. Both sets of data were linked by an identification number for each student that was based on a hash-function of their university email; that is, the researchers only had access to the demographics data in this de-identified form. Gender and ethnicity were included in the demographic data. We removed students whose gender was missing.

For the results related to gender, we report data for students who were enrolled in Physics 1 in the fall 2015 and Physics 2 in the spring 2016 (i.e., students who were enrolled in the "on-cycle" sequence of the courses) to simplify interpretations of changes across the courses. Results are discussed separately for students who were enrolled in calculus-based and algebra-based introductory physics courses, given the differential representation across those divides, the likely differences in haven taken high school physics, and the centrality of physics to their own majors.

Table II shows the percentages of males and females in introductory physics courses who completed the survey at the three points in time. The number of students is different at different points of time due to the fact that some students did not take physics 2 courses and some students may not have been present in the lecture or recitation section in which the survey was administered (and therefore did not participate in the survey).

**Table II.** The total number of students and the percentages of students who completed the survey at different points in time broken down by males (M) and females (F).

|  | Algebra-Based Physics 1 | Algebra-Based Physics 2 | Calculus-Based Physics 1 | Calculus-Based Physics 2 |
|---|---|---|---|---|
| Fall 2015 (N=1,095) | n=421; 64% F, 36% M | n=85; 60% F, 40% M | n=467; 31% F, 69% M | n=122; 28% F, 72% M |
| Beginning of Spring 2016 (N=1,149) | n=306; 59% F, 41% M | n=380; 63% F, 37% M | n=119; 38% F, 62% M | n=344; 34% F, 66% M |
| End of Spring 2016 (N=871) | n=243; 61% F, 39% M | n=295; 63% F, 37% M | n=59; 41% F, 59% M | n=274; 35% F, 65% M |

## IV.    RESULTS

### A.  Gender differences at the beginning of Physics 1

We found that there were significant gender differences in male and female students' motivational factors at the beginning of Physics 1 in both calculus-based and algebra-based courses. We performed a one-way multivariate analysis of variance (MANOVA) with the dependent variables being students' average reported self-efficacy, fascination, value, intelligence mindset, physics epistemology, and grit at the beginning of Physics 1 and the independent variable being gender. For the calculus-based course, the results of the MANOVA revealed that there was a statistically significant difference in male and female students' overall motivation, $F(7,459)=10.344$, $p<0.001$; Wilks' Lambda=0.864. Follow-up univariate tests revealed that males reported statistically significantly higher levels of self-efficacy, fascination, value associated with physics, and physics intelligence mindset. See Table III and Figure 2 for p-values and Cohen's $d$ effect sizes (Cohen's $d$ is $d = (\mu_1 - \mu_2)/\sigma_{pooled}$, where $\mu_1$ is the average score of males and $\mu_2$ is the average score of females and $\sigma_{pooled}$ is the standard deviation of the average score of all students).

For the algebra-based course, the results of the MANOVA revealed that there was a statistically significant difference in male and female students' overall motivation, $F(7,412)=10.213$, $p<0.001$; Wilks' Lambda=0.852. Follow-up univariate tests revealed that males reported statistically significantly higher levels of self-efficacy and fascination. On the other hand, females reported statistically significantly higher levels of grit. See Table III and Figure 3 for exact p-values and effect sizes.

**Table III.** Effect sizes showing the comparison of factors related to males' and females' motivation throughout introductory physics course sequence. Bolded effect size values indicate statistically significant differences at the level of p<0.05.

| | Male > Female effect size | | |
|---|---|---|---|
| **Calculus-Based** | Beginning of Physics 1 | Beginning of Physics 2 | End of Physics 2 |
| Self-Efficacy | **0.75** | **0.80** | **0.95** |
| Fascination | **0.37** | **0.63** | **0.89** |
| Value | **0.32** | **0.35** | **0.70** |
| Physics Intelligence Mindset | **0.22** | **0.43** | **0.28** |
| Physics Epistemology | 0.05 | 0.08 | 0.18 |
| General Intelligence Mindset | -0.09 | -0.01 | -0.12 |
| Grit | -0.17 | -0.17 | -0.02 |
| **Algebra-Based** | | | |
| Self-Efficacy | **0.63** | **0.61** | **0.62** |
| Fascination | **0.60** | **0.83** | **0.56** |
| Value | 0.20 | **0.39** | **0.35** |
| Physics Intelligence Mindset | 0.20 | **0.24** | **0.48** |
| Physics Epistemology | 0.02 | 0.13 | 0.11 |
| General Intelligence Mindset | -0.04 | -0.02 | 0.03 |
| Grit | **-0.26** | -0.20 | **-0.42** |

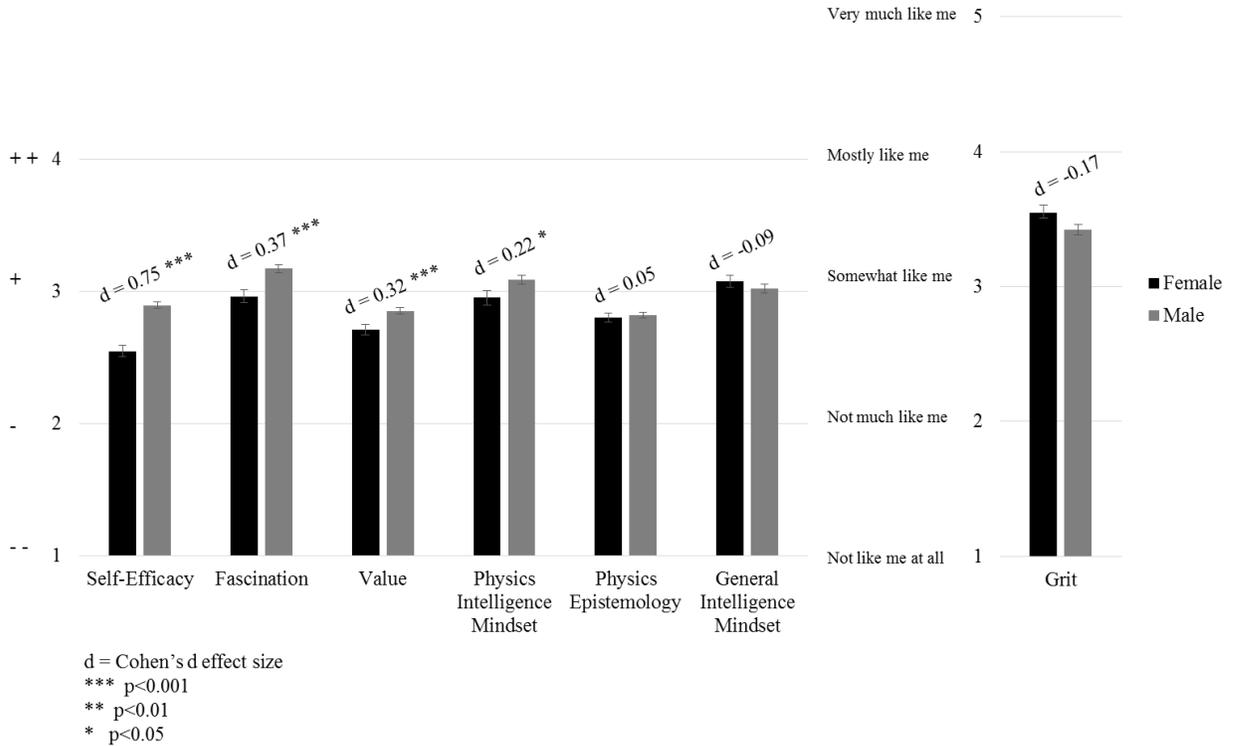

d = Cohen's d effect size
*** p<0.001
** p<0.01
* p<0.05

**Figure 2**. Females' (N=144) and males' (N=323) average reported motivational characteristics at the beginning of calculus-based Physics 1, with standard error bars. The "+" signs indicate positive valence responses and the " – " signs indicate negative valence responses. We separate the construct of grit since it was a 5-point Likert scale.

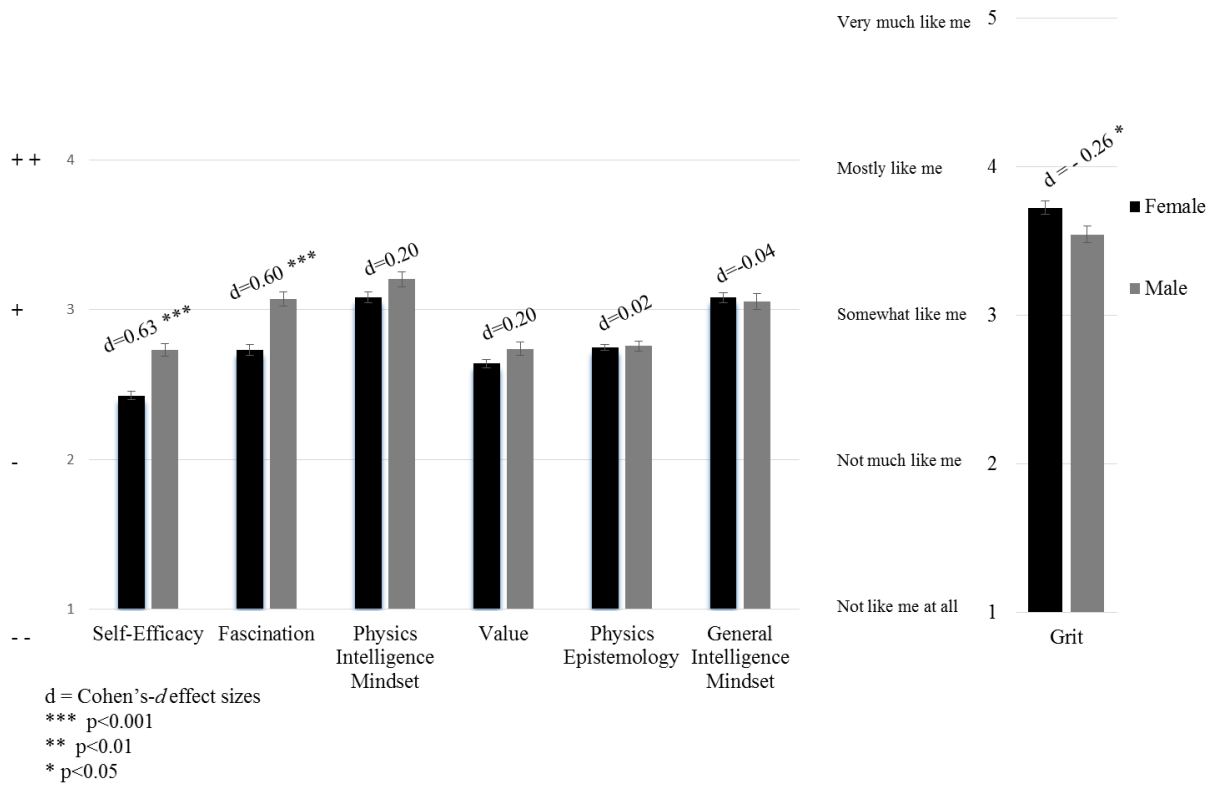

**Figure 3**. Females' (N=270) and males' (N=151) average reported motivational characteristics at the beginning of algebra-based Physics 1, with standard error bars. The "+" signs indicate positive responses and the "-" signs indicate negative responses. We separate the construct of grit since it was a 5-point Likert scale.

**B. Changes in students' motivational characteristics over the course of an introductory physics course sequence**

We calculated change scores for the change in students' motivational characteristics between the beginning of Physics 1 and the beginning of Physics 2. For example, in order to calculate a student's self-efficacy change score in Physics 1, we subtracted the student's average self-efficacy score at the beginning of Physics 1 from the student's average self-efficacy score at the beginning of Physics 2. This change score allowed us to determine the extent to which students' motivational characteristics changed after taking an introductory Physics 1 course focusing on mechanics and Newton's laws. A one-way MANOVA was performed with the dependent variables being the change scores in students' average self-efficacy, fascination, value, physics intelligence mindset, general intelligence mindset, physics epistemology, and grit and the independent variable being gender. We found that there was not a significant difference in the change in male and female students' overall motivation from the beginning of Physics 1 to the beginning of Physics 2 in either calculus-based or algebra-based courses. In other words, on average, the gaps between male and female students' motivational characteristics remained the same. We did find that there was a significant difference in the change scores of males and females in regards to fascination in calculus-based Physics 1 courses (result of ANOVA gives F(1, 280)=5.688, p=0.018). That is, females' average reported fascination with physics decreased significantly more than males' after taking a Physics 1 course and the gap between male and female students' reported fascination increased. See Figure 4 for the change in male and female students' fascination after taking a calculus-based Physics 1 course.

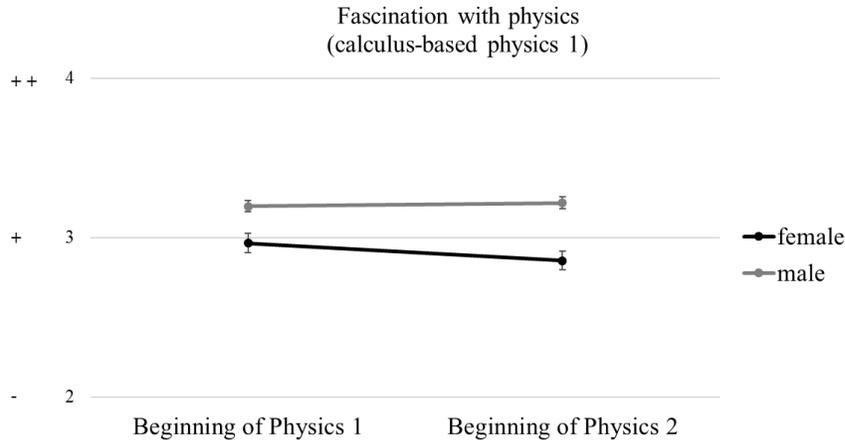

**Figure 4**. Male and female students' fascination with physics before and after taking calculus-based Physics 1, with standard error bars. The "+" signs indicate positive responses and the "-" signs indicate negative responses.

We also calculated change scores for the change in students' motivational characteristics between the beginning of Physics 2 and the end of Physics 2. For example, in order to calculate a student's self-efficacy change score in Physics 2, we subtracted the student's average self-efficacy score at the beginning of Physics 2 from the student's average self-efficacy score at the end of Physics 2. This change score allowed us to determine the extent to which students' motivational characteristics changed after taking an introductory Physics 2 course focusing on electricity and magnetism. A one-way MANOVA was performed with the dependent variables being the change scores in students' average self-efficacy, fascination, value, physics intelligence mindset, general intelligence mindset, physics epistemology, and grit and the independent variable being gender. We found that there was not a significant difference in the change in male and female students' overall motivation from the beginning of Physics 2 to the end of Physics 2 in either calculus-based or algebra-based courses. In other words, the gaps between male and female students' motivational characteristics remained approximately the same after taking a Physics 2 course. However, we did find that there was a significant difference in the change scores of males and females in regards to fascination and value in calculus-based courses. That is, females' average fascination decreased significantly more than males (ANOVA result gives $F(1,246)=6.412$, $p=0.012$) after taking a calculus-based Physics 2 course. Furthermore, females' value associated with physics decreased significantly more than males' (ANOVA result gives $F(1,246)=5.637$, $p=0.018$) in calculus-based Physics 2. In other words, the gap between female and male students' reported fascination and value associated with physics increased after taking a calculus-based Physics 2 course. See Figure 5 for the change in male and female students' fascination and value after taking a calculus-based physics 2 course.

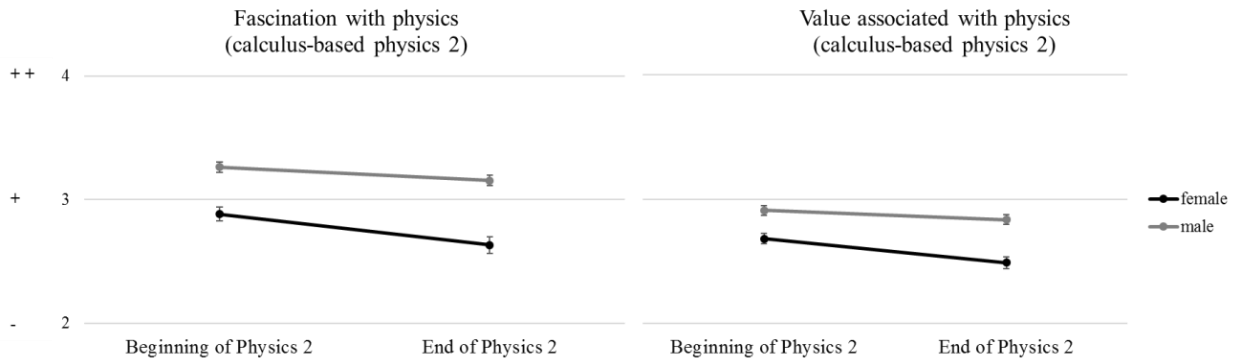

**Figure 5.** Students' fascination and value associated with physics at the beginning and end of calculus-based physics 2, with standard error bars. The "+" signs indicate positive responses and the "-" signs indicate negative responses.

In addition, algebra-based Physics 2, we found that there was a significant difference in the change scores of male and female students in regards to physics intelligence mindset. That is, females' reported physics intelligence mindset became more "fixed" relative to males and the gap between female and male students' physics intelligence mindset increased (ANOVA result gives $F_{(1, 278)}=5.359$, p=0.021). See Figure 6 for the change in male and female students' physics intelligence mindset after taking an algebra-based physics 2 course. See the appendix Figures 1-7 for male and female students' average self-efficacy, fascination, value associated with physics, physics intelligence mindset, physics epistemology, general intelligence mindset, and grit over the course of an introductory physics course sequence (we note that Appendix Figures 1-7 include all students who responded to the survey, not just the common cohort of students taking both Physics 1 and Physics 2).

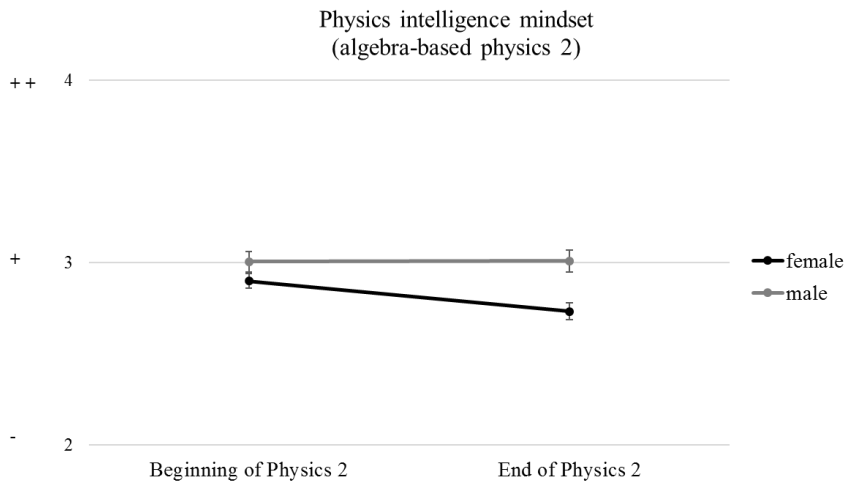

**Figure 6**. Students' physics intelligence mindset at the beginning and end of algebra-based physics 2, with standard error bars. The "+" signs indicate positive responses (i.e., a more malleable view of intelligence) and the "-" signs indicate negative responses (i.e., a more fixed view of intelligence).

## V. SUMMARY AND DISCUSSION

We found gender differences with regards to students' views of several motivational factors throughout a two-semester introductory physics courses. At the beginning of a Physics 1 course, females' average self-efficacy, fascination, value associated with physics, and physics intelligence mindset were significantly lower than males'. We also found that females' fascination and value associated with physics decreased significantly more than males' after taking a calculus-based physics course sequence. Furthermore, after taking an algebra-based physics 2 course focusing on electricity and magnetism, females' physics intelligence mindset became more "fixed" as compared to males' physics intelligence mindset. This result suggests that women may view intelligence in physics differently than intelligence in general, e.g., that one must be "brilliant" to do physics. We found that grit was the only construct on which females reported average scores higher than males.

We discussed a holistic framework, SELF, that can be used as a guide to increase students' engagement with introductory physics courses. The SELF framework posits that instructional design, learning tools and their implementation along with student characteristics and environments play a critical role in helping students learn effectively. Since student characteristics can play a central role in whether students are engaged in learning effectively, the baseline data we discussed for students' characteristics can be a useful stepping stone for developing and implementing pedagogies and learning tools to help all students succeed in introductory physics courses.

## VI. IMPLICATIONS

Students' characteristics discussed here can play a central role in students' engagement with a course and impact their learning. We found important gender differences on motivational factors such as value, fascination, self-efficacy, and intelligence mindset. Researchers, curriculum developers, and instructors can take these differences into account when developing and implementing instruction and learning tools to help all students engage effectively with physics courses.

Our findings suggest that that females who were interested in physics initially (had high value and fascination associated with physics at the beginning of a physics course) lose interest over the introductory physics course sequence. Instructors, researchers, and curriculum developers may be able to boost females' engagement with physics courses by framing physics as a science that can benefit society throughout a physics course sequence. Indeed, a recent study showed that when a STEM career was presented to females as more altruistic, communal, and beneficial to people, females' interest in STEM fields increased [30,102]. Framing instruction and learning tools appropriately to get buy-in from students (factor 3 in the SELF framework), e.g., framing physics as a science that can benefit society, is one way to motivate all students in introductory physics courses to be engaged with curriculum and learning tools effectively.

In addition, we found that women tend to enter physics courses with lower self-efficacy than men, and their self-efficacy remains low after taking introductory physics courses. Low self-efficacy may be associated with women feeling like they do not belong and that they are not capable of doing physics. Certain types of instruction and learning tools in Factor 1 in the SELF framework (when implemented appropriately) can improve students' self-efficacy. For example, research has shown that instructional strategies such as collaborative learning, conceptual problems, and inquiry-based labs contributed to students' improved self-efficacy and improved classroom climate [40]. Modeling Instruction has also been shown to improve students' self-efficacy, especially for females [11]. Furthermore, the implementation of instruction and learning tools (Factor 3 of the SELF framework) can impact students' self-efficacy. For example, research suggests that when implementing collaborative group work, collaborative groups should include at least two underrepresented students in a group because an underrepresented student's ideas and contributions are sometimes ignored if he/she is the only underrepresented student in the group [103]. The timing of instructional interventions can also be taken into account—since women tend to have lower self-efficacy at the beginning of a physics course sequence, this issue should be addressed early on in the course.

Our research also suggests that women tended to view intelligence in physics as a fixed ability, compared to men. This type of mindset can affect learning. For example, Dweck found that students who believe that intelligence is fixed are more likely to attribute academic setbacks to a lack of ability than students who believe that intelligence is malleable and improvable with hard work and effort [53]. These views shape how students respond to setbacks, either by withdrawing effort or redoubling effort, seeking help, using a better strategy, etc. It may be beneficial to provide explicit support to students to improve their views about intelligence during the implementation of learning tools (Factor 3 of the SELF framework), such as discussing the importance of using effective learning strategies, attributing failure to inappropriate learning strategies as opposed to intellectual ability, and viewing mistakes as an opportunity for learning and growth throughout an introductory physics course sequence. Indeed, interventions have been developed that help students view intelligence as malleable. For example, Blackwell and colleagues designed an intervention for middle school students in which they attended an eight-session workshop to learn about study skills and scientific research showing that the brain grows connections and "gets smarter" when a person works on challenging tasks. Students who learned that intelligence is malleable earned better math grades over the course of the year than students in the control group [54]. Other interventions involved providing effort v. intelligence praise after success [104] and orchestrating email exchanges with a mentor during a yearlong program [105].

Instructors, researchers, and curriculum developers can create and implement effective instruction and learning tools in physics, in part, by considering students' motivational characteristics. Focusing on students' motivational characteristics, especially those of underrepresented students, may prove fruitful in helping more students succeed in STEM courses and increase the diversity in STEM fields.

## ACKNOWLEDGEMENTS

We thank the United States National Science Foundation for award DUE-1524575.

## APPENDIX

## Final version of the survey of factors related to students' motivation

1. I wonder about how nature works
   a. Never
   b. Once a month
   c. Once a week
   d. Every day
2. In general, I find physics
   a. Very boring
   b. boring
   c. interesting
   d. Very interesting
3. Knowing physics helps me understand how the world works:
   a. Never
   b. Sometimes
   c. Most of the time
   d. All of the time
4. Knowing physics is important for:
   a. No jobs
   b. A few jobs
   c. Most jobs
   d. All jobs
5. I can complete the physics activities I get in a lab class
   a. Rarely
   b. Half the time
   c. Most of the time
   d. All the time
6. If I went to a museum, I could figure out what is being shown about physics in:
   a. None of it
   b. A few areas
   c. Most areas
   d. All areas

| | No! | no | yes | Yes! |
|---|---|---|---|---|
| 7. I want to know everything I can about physics | a | b | c | d |
| 8. Physics makes the world a better place to live. | a | b | c | d |
| 9. Knowing physics is important for being a good citizen. | a | b | c | d |
| 10. Learning physics helps me understand situations in my everyday life. | a | b | c | d |
| 11. I am often able to help my classmates with physics in the laboratory or in recitation. | a | b | c | d |
| 12. I get a sinking feeling when I think of trying to tackle difficult physics problems. | a | b | c | d |

| | | Strongly disagree | Disagree | Agree | Strongly agree |
|---|---|---|---|---|---|
| 15. | You have a certain amount of intelligence, and you can't really do much to change it. | a | b | c | d |
| 16. | No matter how much intelligence you have, you can always change it quite a bit. | a | b | c | d |
| 17. | Anyone can become good at solving physics problems through hard work. | a | b | c | d |
| 18. | Only very few specially qualified people are capable of really understanding physics. | a | b | c | d |

| | | Not like me at all | Not much like me | Somewhat like me | Mostly like me | Very much like me |
|---|---|---|---|---|---|---|
| 19. | I often set a goal but later choose to pursue a different one. | a | b | c | d | e |
| 20. | I have difficulty maintaining my focus on projects that take more than a few months to complete. | a | b | c | d | e |
| 21. | I finish whatever I begin. | a | b | c | d | e |

| | | | | Strongly disagree | Disagree | Agree | Strongly agree |
|---|---|---|---|---|---|---|---|
| 13. | If I wanted to, I could be good at doing physics research. | | | a | b | c | d |
| 14. | If I study, I will do well on a physics test. | | | a | b | c | d |

| | | Strongly disagree | Disagree | Agree | Strongly agree |
|---|---|---|---|---|---|
| 22. | I do not expect to understand physics equations in an intuitive sense; they must just be taken as givens. | a | b | c | d |
| 23. | Knowledge in physics consists of many pieces of information each of which applies primarily to a specific situation. | a | b | c | d |
| 24. | Learning physics is mainly about remembering the laws, principles, and equations given in class and/or in the textbook. | a | b | c | d |
| 25. | Problem solving in physics mainly involves matching problems with facts or equations and then substituting the values to get a number. | a | b | c | d |
| 26. | In doing a physics problem, if my calculation gives a result that differs significantly from what I expect, I'd have to trust the calculation. | a | b | c | d |
| 27. | "Understanding" physics basically means being able to recall something you've read or been shown. | a | b | c | d |
| 28. | When doing an experiment, I try to understand how the experimental setup works. | a | b | c | d |
| 29. | The primary purpose of doing a physics experiment is to confirm previously known results. | a | b | c | d |

**Factor loadings for each question on the survey**

| Rotated Component Matrix[a] | | | | | | |
|---|---|---|---|---|---|---|
| | Component | | | | | |
| | 1 | 2 | 3 | 4 | 5 | 6 |
| I wonder about how nature works | | .276 | | .337 | | |
| In general, I find physics | .383 | .456 | | .444 | | |
| Knowing physics helps me understand how the world works: | .221 | .562 | | .350 | | |
| Knowing physics is important for: | | .641 | | | | |
| I can complete the physics activities I get in a lab class | .678 | | | | | |
| If I went to a museum, I could figure out what is being shown about physics in... | .572 | .277 | | | | |
| I want to know everything I can about physics | .302 | .552 | | .423 | | |
| Physics makes the world a better place to live | .244 | .496 | | .228 | | |
| Knowing physics is important for being a good citizen | | .719 | | -.210 | | |
| Learning physics helps me understand situations in everyday life | | .661 | | | | |
| I am often able to help my classmates with physics in the laboratory or in recitation | .743 | | | | | |
| I get a sinking feeling when I think of trying to tackle difficult physics problems | .595 | | .226 | .286 | | |
| If I wanted to, I could be good at doing physics research | .582 | .240 | | .322 | | |
| If I study, I will do well on a physics test | .509 | | | .336 | .230 | |
| You have a certain amount of intelligence, and you can't really do much to change it | | | | | .825 | |
| No matter how much intelligence you have, you can always change it quite a bit | | | | | .814 | |
| Anyone can become good at solving physics problems through hard work | | | | .411 | .511 | |
| Only very few specially qualified people are capable of really understanding physics | | | .210 | .306 | .444 | |
| I often set a goal but later choose to pursue a different one | | | | | | .764 |
| I have difficulty maintaining my focus on projects that take more than a few months to complete | | | | | | .772 |
| I finish whatever I begin | | | | | | .743 |
| I do not expect to understand physics equations in an intuitive sense; they must be taken as givens | .311 | | .405 | .358 | | |
| Knowledge in physics consists of many pieces of information each of which applies primarily to a specific situation | | | .584 | | | |
| Learning physics is mainly about remembering the laws, principles, and equations given in class and/or in the textbook | | | .666 | | | |
| Problem solving in physics mainly involves matching problems with facts or equations and then substituting the values to get a number | | | .646 | | | |
| In doing a physics problem, if my calculation gives a result that differs significantly from what I expect, I'd have to trust the calculation | | | .392 | .332 | | |
| Understanding physics basically means being able to recall something you've read or been shown. | | | .671 | | | |
| When doing an experiment, I try to understand how the experimental setup works | | | | .563 | | |
| The primary purpose of doing a physics experiment is to confirm previously known results | | | .547 | | | |
| Extraction Method: Principal Component Analysis. Rotation Method: Varimax with Kaiser Normalization. | | | | | | |
| a. Rotation converged in 10 iterations. | | | | | | |

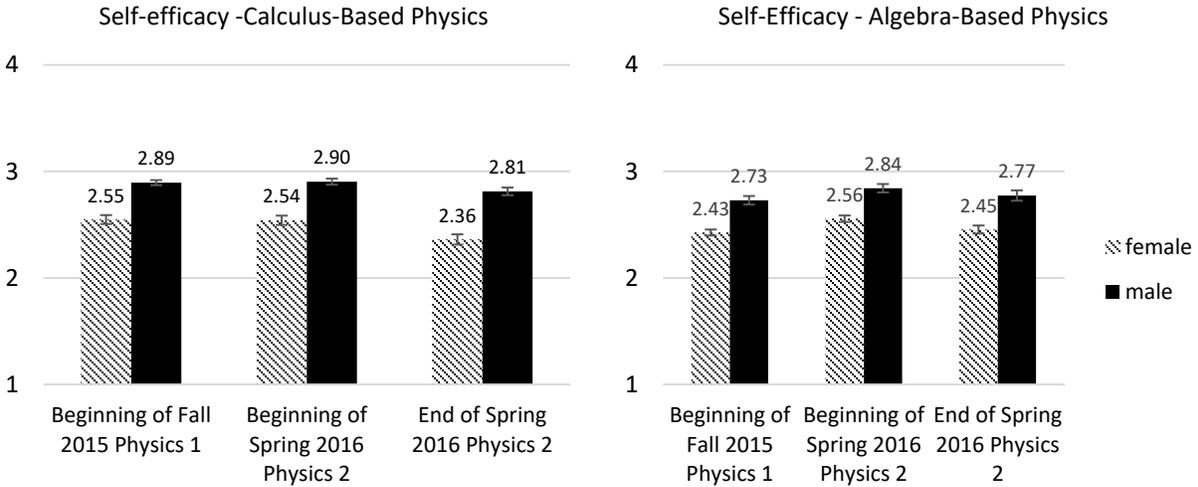

**Figure 1.** Female and male students' average self-efficacy scores throughout an introductory physics course sequence for calculus-based courses (left) and algebra-based courses (right) with standard error bars.

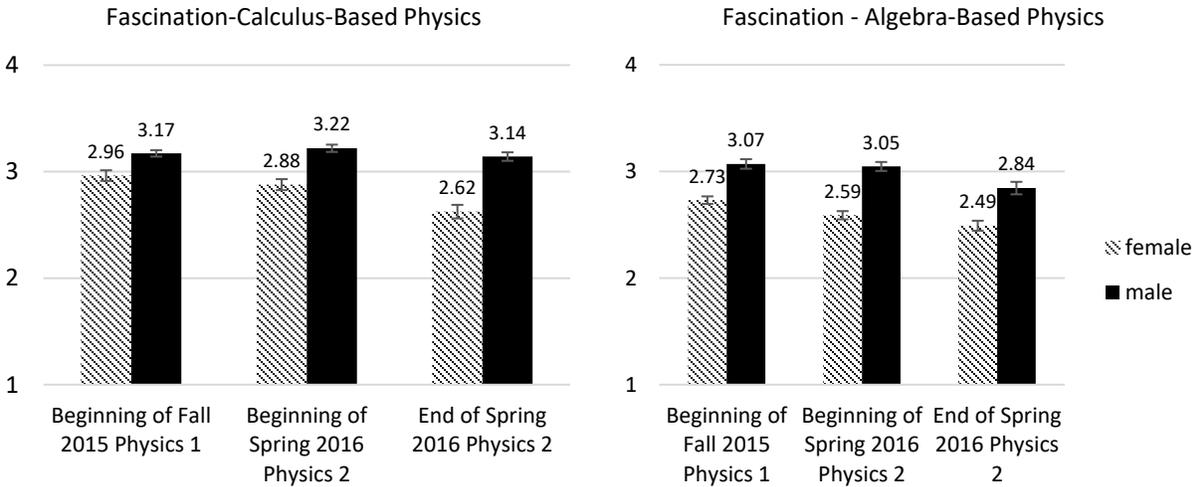

**Figure 2.** Female and male students' average fascination scores throughout an introductory physics course sequence for calculus-based courses (left) and algebra-based courses (right) with standard error bars.

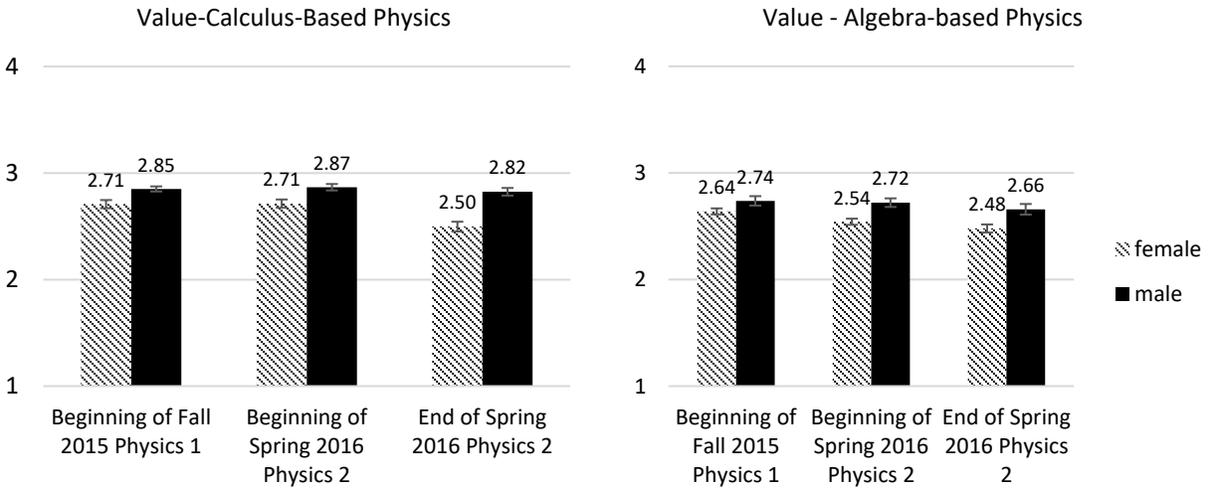

**Figure 3.** Female and male students' average value scores throughout an introductory physics course sequence for calculus-based courses (left) and algebra-based courses (right) with standard error bars.

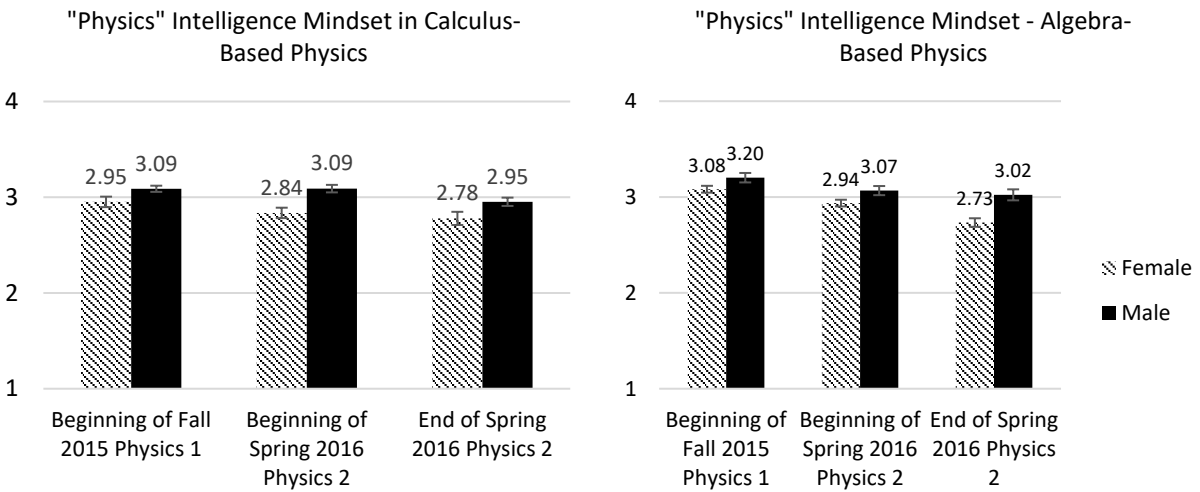

**Figure 4.** Female and male students' average "physics" intelligence mindset scores throughout an introductory physics course sequence for calculus-based courses (left) and algebra-based courses (right) with standard error bars.

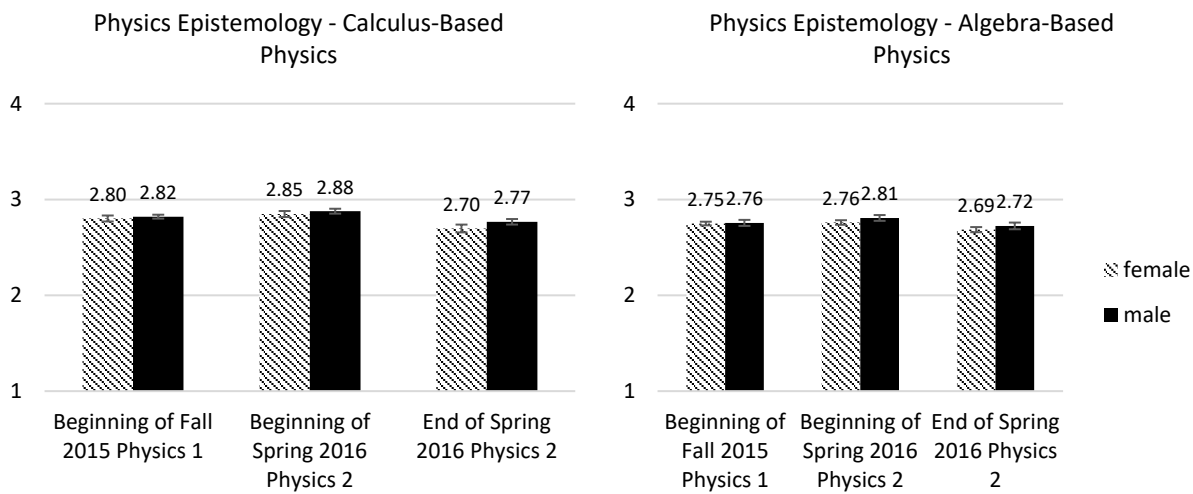

**Figure 5.** Female and male students' average physics epistemology scores throughout an introductory physics course sequence for calculus-based courses (left) and algebra-based courses (right) with standard error bars.

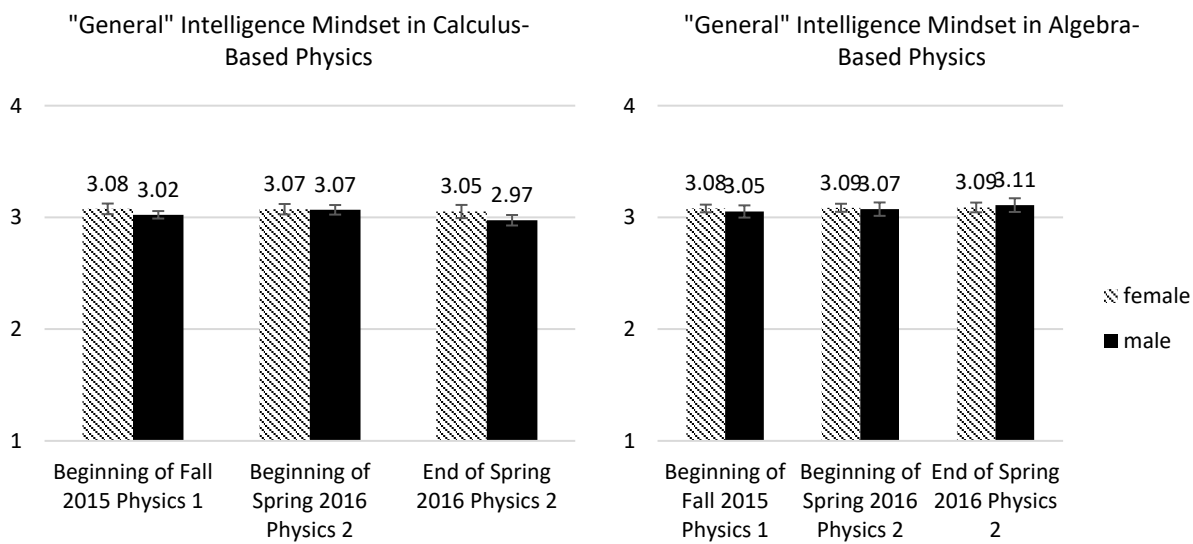

**Figure 6.** Female and male students' average intelligence mindset scores throughout an introductory physics course sequence for calculus-based courses (left) and algebra-based courses (right) with standard error bars.

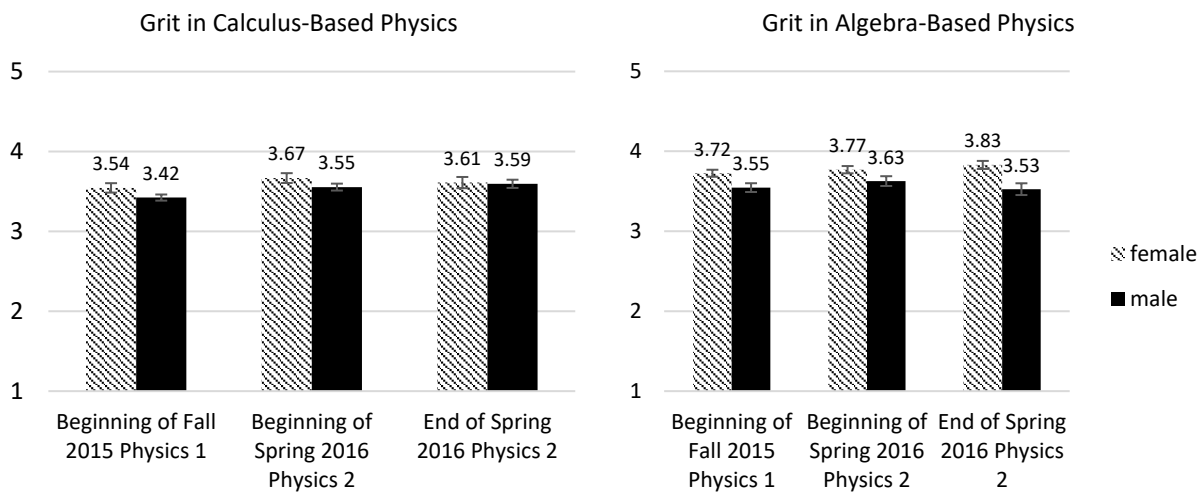

**Figure 7.** Female and male students' average grit scores throughout an introductory physics course sequence for calculus-based courses (left) and algebra-based courses (right) with standard error bars.